\documentclass[12pt]{iopart}

\usepackage{graphicx}
\usepackage{iopams}
\begin{document}

\title{A Numerical Study of Coulomb Interaction Effects on 2D Hopping Transport}

\author{Yusuf A. Kinkhabwala\dag, Viktor A. Sverdlov\dag\ and Konstantin K. Likharev\dag}
\address{\dag\ Department of Physics and Astronomy, Stony Brook
University, Stony Brook, NY 11794-3800}

\begin{abstract}
We have extended our supercomputer-enabled Monte Carlo simulations
of hopping transport in completely disordered 2D conductors to the
case of substantial electron-electron Coulomb interaction. Such
interaction may not only suppress the average value of hopping
current, but also affect its fluctuations rather substantially. In
particular, the spectral density $S_I \left( f \right)$ of current
fluctuations exhibits, at sufficiently low frequencies, a $1/f$-like
increase which approximately follows the Hooge scaling, even at
vanishing temperature. At higher $f$, there is a crossover to a
broad range of frequencies in which $S_I \left( f\right)$ is nearly
constant, hence allowing characterization of the current noise by
the effective Fano factor $F\equiv S_I\left( f \right)/2e
\left\langle I\right\rangle$. For sufficiently large conductor
samples and low temperatures, the Fano factor is suppressed below
the Schottky value ($F=1$), scaling with the length $L$ of the
conductor as $F = \left( L_c / L \right)^{\alpha}$. The exponent
$\alpha$ is significantly affected by the Coulomb interaction
effects, changing from $\alpha = 0.76 \pm 0.08$ when such effects
are negligible to virtually unity when they are substantial. The
scaling parameter $L_c$, interpreted as the average percolation
cluster length along the electric field direction, scales as $L_c
\propto E^{-(0.98 \pm 0.08)}$ when Coulomb interaction effects are
negligible and $L_c \propto E^{ -(1.26 \pm 0.15)}$ when such effects
are substantial, in good agreement with estimates based on the
theory of directed percolation.
\end{abstract}

%Uncomment for PACS numbers title message

\maketitle

\section{\label{sec:level1}Introduction}

The hopping transport of quasi-localized electrons in disordered
conductors and semiconductors has been studied for many years - for
comprehensive reviews, see
Refs.~\cite{MottBook,ShklovskiiBook,EfrosPollackCollection}. The
more recent observation
\cite{AverinLikharev1991,MatsuokaLikharev1998} that hopping
transport may implement the quasi-continuous (``sub-electron")
charge transfer, hence providing a possible solution to the random
background charge problem in single-electronics \cite{Likharev1999},
has renewed interest in this phenomenon, with an emphasis on the
shot noise of the hopping current \cite{1DKorotkovLikharev2000,
Kuznetsovetal2000, 2DSverdlovKorotkovLikharev2001,
2DKinkhabwalaSverdlovKorotkovLikharev2004}. The objective of this
paper is to present the results of an extension of our previous
numerical studies of 2D hopping
\cite{2DSverdlovKorotkovLikharev2001,
2DKinkhabwalaSverdlovKorotkovLikharev2004} to the case of
substantial Coulomb interaction of hopping electrons. Just as in the
case of negligible interaction
\cite{2DKinkhabwalaSverdlovKorotkovLikharev2004}, the use of
advanced algorithms of spectral density calculation
\cite{Method-KinkhabwalaSverdlovKorotkov2005} and modern
supercomputer facilities has allowed us not only to obtain more
complete and exact results for average characteristics of hopping
transport (including the dependence of the dc current on temperature
and electric field), but also to calculate the spectral density of
current fluctuations at low temperatures.

In order to explain our new findings, we have to start with a brief
summary of the basic prior results.

\subsection{\label{sec:level1} Coulomb Gap}

Most theoretical discussions of the Coulomb interaction effects on
hopping are based on the notion of the Coulomb gap in the electron
energy spectrum. Generally speaking, substantial Coulomb interaction
makes the single-particle energy meaningless. However, the
introduction \cite{ShklovskiiBook} of the \textit{effective}
single-particle energy $\varepsilon$, which includes the
contribution from the Coulomb interaction with other electrons,
immediately leads to a ``soft" gap in the single-particle density of
states $\nu\left(\varepsilon \right)$ at $\varepsilon \approx \mu$,
where $\mu$ is the Fermi level. In the case of 2D conductors with
the 3D Coulomb interaction law, which is the focus of our current
work, simple arguments \cite{ShklovskiiBook,EfrosPollackCollection}
yield
\begin{equation}
\nu\left(\varepsilon \right) = c \frac{\kappa^2}{e^4} \left\vert
\varepsilon - \mu \right\vert, \label{eq:singleparticleDoS}
\end{equation}
where $e$ is electron charge, $\kappa$ is the dielectric constant of
the insulating environment and $c$ is a dimensionless constant.
Equation~(\ref{eq:singleparticleDoS}) is valid only when the 2D
density of states  $\nu\left(\varepsilon \right)$ is much smaller
than the ``seed" density of states $\nu_0$; for larger $\varepsilon$
there is a continuous crossover to $\nu_0$. The effective width
$\Delta$ of the Coulomb gap can be estimated from the natural
condition $\nu\left(\Delta \right) = \nu_0$, resulting in
\begin{equation}
\Delta = \frac{e^4 \nu_0}{c \kappa^2}. \label{eq:CoulombGapWidth}
\end{equation}
A self-consistent approach  \cite{EfrosPollackCollection} allows a
more rigorous evaluation of the Coulomb gap width, giving $c =
2/\pi$.

\subsection{\label{sec:level1} DC Transport Characteristics}

At low applied electric fields $E$, the average current
$\left\langle I\right\rangle$ is a linear function of $E$, i.e$.$
the 2D (``sheet") dc conductivity $\sigma \left(T, E, \chi \right)
\equiv \left\langle I\right\rangle/E W$ (where $W$ is the width of
the conductor) is independent of $E$. For not very high temperatures
($T \ll T_0$, where $k_B T_0 \equiv 1/\nu_0 a^2$  and $a$ is the
localization radius), the ratio $\sigma/\sigma_0$ (where $\sigma_0$
is a constant characterizing the sample) depends only on two
dimensionless parameters: ratio $T/T_0$ and parameter $\chi \equiv
e^2 \nu_0 a/ \kappa$ characterizing the Coulomb interaction
strength. The relation between these two parameters determines two
possible variable-range hopping transport regimes.

If the Coulomb interaction is weak ($\chi^3 \ll T/T_0$), the average
length $r\left(T, E, \chi \right)$ of the so-called ``critical
hops", which connect percolation clusters and hence determine the
current, may be found from the Mott theory \cite{MottBook,
ShklovskiiBook, EfrosPollackCollection}:
\begin{equation}
r\left(T, 0, 0 \right) \propto \left( \frac{T_0}{T} \right)^{1/3} a.
\label{eq:zerocoulombhightemperaturehoplength}
\end{equation}
In this case the conductivity is
\cite{MottBook,ShklovskiiBook,EfrosPollackCollection}:
\begin{equation}
\frac{\sigma}{\sigma_0} \approx A\left(T,0,0 \right)\exp
\left[-\left( B\left( T,0,0 \right) \frac{T_0}{T}
\right)^{1/3}\right],
\label{eq:zerocoulombhightemperatureconductivity}
\end{equation}
where $A\left(T, E, \chi \right)$ is a dimensionless,
model-dependent, slow function of its arguments, while $B\left(T, E,
\chi \right)$ is usually treated as a constant, but in general may
be also a weakly dependent function of its arguments.

On the other hand, if the Coulomb interaction is strong ($\chi^3 \gg
T/T_0$), the critical hops are longer
\cite{ShklovskiiBook,EfrosPollackCollection}:
\begin{equation}
r\left( T, 0, \chi \right) \propto \left( \frac{\chi T_0}{T}
\right)^{1/2} a, \label{eq:nonzerocoulombhightemperaturehoplength}
\end{equation}
and the dc conductivity is suppressed
\cite{ShklovskiiBook,EfrosPollackCollection}:
\begin{equation}
\frac{\sigma}{\sigma_0} \approx A\left(T,0,\chi \right) \exp
\left[-\left( B\left(T,0,\chi \right) \frac{\chi T_0}{T}
\right)^{1/2}\right].
\label{eq:nonzerocoulombhightemperatureconductivity}
\end{equation}

In the case of relatively high electric fields ($E \gg E_T$, where
$E_T \equiv k_B T /e a$), the dc current is a highly nonlinear
(exponential) function of the applied electric field $E$. If the
field is not extremely high ($E \ll E_0 \equiv 1/e \nu_0 a^3$), i.e.
in the variable-range hopping domain, we can again distinguish two
different transport regimes.

If the Coulomb interaction is weak ($\chi^3 \ll E/E_0$), one can
neglect the effects of Coulomb interaction to evaluate the critical
hop length
\begin{equation}
r\left(0, E, 0 \right) \propto \left( \frac{E_0}{E} \right)^{1/3} a.
\label{eq:zerocoulombhighelectricfieldhoplength}
\end{equation}
In this case, the dc conductivity is
\cite{Shklovskii1973,ApsleyHughes19741975,PollackRiess1976,RentzschShlimakBerger1979,vanderMeerSchuchardtKeiper1982}
\begin{equation}
\frac{\sigma}{\sigma_0} \approx A \left( 0, E, 0 \right) \exp
\left[-\left( B \left( 0, E, 0 \right) \frac{E_0}{E} \right)^{1/3}
\right]. \label{eq:zerocoulombhighelectricfieldconductivity}
\end{equation}

In the opposite limit ($\chi^3 \gg E/E_0$),
\begin{equation}
r\left(0, E, \chi \right) \propto \left( \frac{\chi E_0}{E}
\right)^{1/2} a, \label{eq:nonzerocoulombhighelectricfieldhoplength}
\end{equation}
and the dc conductivity is lower \cite{RentzschShlimakBerger1979}:
\begin{equation}
\frac{\sigma}{\sigma_0} \approx A \left( 0, E, \chi \right) \exp
\left[-\left( B \left( 0, E, \chi \right) \frac{\chi E_0}{E}
\right)^{1/2} \right].
\label{eq:nonzerocoulombhighelectricfieldconductivity}
\end{equation}

\subsection{\label{sec:level1}Current Fluctuations}

At low temperatures \cite{thermalnoise}, the dynamical fluctuations
of the current flowing through a mesoscopic system are more
sensitive to the charge transport mechanism peculiarities than the
average transport characteristics, and therefore may provide
additional information about the conduction physics \cite{KoganBook,
JongBeenakker1997, BlanterButtiker2000}. If we refrain from the
discussion of the quantum fluctuations at extremely high
frequencies, two basic frequency ranges have to be distinguished. At
very low frequencies, one can expect the $1/f$-type noise that is
observed experimentally in a wide variety of conductors - see, e.g.,
Ref.~\cite{KoganBook}. In most cases the noise scales approximately
in accordance with the phenomenological Hooge formula
\cite{KoganBook,Hooge1969}. For a 2D conductor, this formula can be
presented as
\begin{equation}
\frac{S_I \left(f \right)}{\left\langle I\right\rangle^2} =
\frac{a^2} {LW} \frac{C\left(f \right)}{f},
\label{eq:hoogenoisescaling}
\end{equation}
where $S_I(f)$ is the current spectral density, $L$ is the length of
the conductor (along the current flow) and $C\left(f \right)$ is
either a dimensionless constant or a weak function of the
observation frequency $f$. In particular, many studies
\cite{KoganBook} have found that $C\left(f \right)/f \propto 1/f^p$,
where $p$ is typically between $1$ and $2$. For the particular case
of hopping conduction, two major theories of $1/f$ noise have been
suggested, based, respectively, on ``carrier number" fluctuations
\cite{Shklovskii1980_1981, Shklovskii2003,ShtengelYu2003} and
``mobility" fluctuations
\cite{Kozub1996,KozubBaranovskiiShlimak1999} as possible origins of
the noise. Unfortunately, both theories have been developed for the
case of substantially nonvanishing temperatures, for which an
accurate numerical study of noise is difficult even with currently
available advanced simulation algorithms and supercomputer
resources.

At relatively high frequencies, the noise spectral density is a very
slow (practically constant) function of $f$, and may be considered
as a mixture of thermal fluctuations and shot noise. In the most
interesting case of sufficiently low temperatures, the thermal
fluctuations are negligible, and the broadband fluctuations are
completely due to electric charge discreteness (shot noise).

An emphasis of most recent studies has been on the suppression of
the shot noise with respect to its Schottky value $2e\left\langle
I\right\rangle$. In particular, such suppression is a necessary
condition for quasi-continuous charge transfer at relatively high
frequencies \cite{AverinLikharev1991, MatsuokaLikharev1998}. If the
current spectral density $S_I(f)$ is flat at $f \rightarrow 0$, it
may be characterized by the Fano factor
\begin{equation}
F\equiv \frac{S_I(0)}{2e\left\langle I\right\rangle},
\label{eq:fanofactor}
\end{equation}
so that the term ``shot noise suppression" means that $F < 1$.
Previous theoretical studies of shot noise at hopping in artificial
(space-ordered) 1D \cite{Derrida1998, 1DKorotkovLikharev2000} and
(both space-ordered and random) 2D
\cite{2DSverdlovKorotkovLikharev2001,
2DKinkhabwalaSverdlovKorotkovLikharev2004} systems have shown that
the shot noise may be, indeed, suppressed, obeying
\begin{equation}
F = \left( \frac{L_c}{L}\right)^{\alpha},\qquad L\gg L_c,
\label{eq:fanolength}
\end{equation}
where $L_c$ is a scaling constant interpreted as the average
percolation cluster length (i.e$.$ the average distance separating
critical hops \cite{ShklovskiiBook, StaufferAharonyBook}) and
$\alpha$ is a positive exponent. In fact, at $T \rightarrow 0$ in
the limit of negligible Coulomb interactions, our prior results
\cite{2DKinkhabwalaSverdlovKorotkovLikharev2004} show that $L_c$
obeys the law
\begin{equation}
L_c = J \left( \frac{E_0}{E} \right)^{\mu}a,
\label{eq:clusterlengthstrongfield}
\end{equation}
where $J$ is a dimensionless constant of the order of $1$, and the
value of the numerical exponent is $\mu=0.98\pm 0.08$, consistent
with the estimate $\mu \approx 0.91$ based on directed percolation
theory \cite{2DKinkhabwalaSverdlovKorotkovLikharev2004,
StaufferAharonyBook, Obukhov1980, Essametal1986}.

Considering a very long conductor, one might suspect that the
electron motion in distant parts should not be correlated. This
assumption immediately leads to $\alpha = 1$
\cite{BlanterButtiker2000}. However, both analytical and numerical
results \cite{1DKorotkovLikharev2000,Derrida1998} show that at 1D
hopping without Coulomb interaction, $\alpha$ may be as low as 1/2.
This nontrivial result may be interpreted as a consequence of an
essentially infinite correlation length in 1D conductors, due to the
on-site interaction of hopping electrons. Even more surprisingly,
the exponent $\alpha$ may be substantially below 1 even in 2D
conductors. For systems on a regular lattice, and without the
Coulomb interaction, numerical modeling
\cite{2DSverdlovKorotkovLikharev2001} yields $\alpha=0.85\pm 0.02$.
In our recent work \cite{2DKinkhabwalaSverdlovKorotkovLikharev2004},
this finding has been confirmed for 2D hopping in conductors with
completely random distribution of localized sites both in space and
in energy. Our most accurate result was $\alpha=0.76\pm 0.08$,
i.e$.$ significantly below 1.

It has not been immediately clear how the inclusion of Coulomb
interaction effects might affect this result. For 1D hopping with
increasing strength of the Coulomb interaction, numerical results
\cite{1DKorotkovLikharev2000} show $\alpha$ crossing over from
nearly 1/2 up to 1. The similar behavior might be expected for 2D
hopping, because the long-range correlations, apparently responsible
for the difference between $\alpha$ and $1$, should be suppressed by
Coulomb interaction effects, provided that the conductor length $L$
is larger than a certain crossover length determined by the
interaction constant $\chi$. Unfortunately, recent experiments
\cite{Kuznetsovetal2000,Roshkoetal2002,Savchenkoetal2004} could not
help in answering this question; while giving a reliable
confirmation of the shot noise suppression in longer conductors,
their accuracy is not sufficient to resolve a possible (relatively
minor) deviation of $\alpha$ from 1.

The resolution of the problem of shot noise suppression in long
conductors has been the main motivation for the numerical experiment
described in this paper. However, since the calculation of dc
transport characteristics is computationally much less demanding
that that of current noise, we have used this opportunity to obtain
accurate values for the slow functions $A$ and $B$ for the same
model of 2D hopping.

\section{\label{sec:level1}Model}

We have studied broad 2D rectangular conductors ($W\gg L_c$) with
``open" boundary conditions on the interfaces with well-conducting
electrodes \cite{2DSverdlovKorotkovLikharev2001,
2DKinkhabwalaSverdlovKorotkovLikharev2004}. The conductors are
assumed to be ``fully frustrated", with a large number of localized
sites randomly distributed over the conductor area. At the sites,
the corresponding electron ``seed" eigenenergies
$\varepsilon^{\left( 0 \right)}$ are also random, being uniformly
distributed over a sufficiently broad energy band $2B$, so that the
2D density of states $\nu_0$ is constant at all energies relevant
for conduction.

The carriers are permitted to hop from any site $j$ to any other
site $k$ with the rate
\begin{equation}
\gamma_{jk}=\Gamma_{jk}\exp \left(-\frac{r_{jk}}{a}\right),
\label{eq:distancerateequation}
\end{equation}
where $r_{jk}$ is the site separation distance and $\Gamma_{jk}$
contains the energy dependence (see below). Such exponential
dependence on the length of a hop is standard for virtually all
theoretical studies of hopping \cite{loclengthfootnote}. Following
our prior work \cite{2DSverdlovKorotkovLikharev2001,
2DKinkhabwalaSverdlovKorotkovLikharev2004}, we take
Eq.~(\ref{eq:distancerateequation}) literally even at small
distances $r_{jk}\sim a$. The energy dependence of $\Gamma_{jk}$ is
given by the usual formula
\cite{2DKinkhabwalaSverdlovKorotkovLikharev2004}
\begin{equation}
\hbar \Gamma_{jk}\left( \Delta U_{jk} \right)=g\frac{\Delta
U_{jk}}{1-\exp\left( -\Delta U_{jk}/k_{B}T\right) },
\label{eq:energyrateequation}
\end{equation}
where $g$ is a dimensionless parameter which determines the 2D
conductivity scale $\sigma_0 \equiv g e^2/\hbar$ \cite{g}, while
$\Delta U_{jk}$ is the difference of the total system energy before
and after the hop from site $j$ to site $k$:
\begin{equation}
\Delta U_{jk} \equiv U_{j} -U_k +e\mathbf{E}\mathbf{r}_{jk}.
\label{eq:systemenergydifference}
\end{equation}
Here $U$ is the total internal energy of the system, including the
effects of Coulomb interaction:
\begin{equation}
U \equiv \sum_{l}\left[n_l\varepsilon^{\left( 0 \right)}_l +
\frac{e^2}{2 \kappa} \left(n_l-\frac{1}{2}\right) \sum_{l'\neq l}
\left(n_{l'}-\frac{1}{2}\right) G\left( r_l, r_{l'}\right)\right],
\label{eq:totalsystemenergy}
\end{equation}
where $n_l$ = 0 or 1 is the occupation number of the $l$-th
localized site. (Similar to earlier studies
\cite{ShklovskiiBook,EfrosPollackCollection} of the Coulomb effect
on hopping, we keep the system electroneutral by adding a background
charge of $e/2$ to each site.) $G\left( r_j, r_k\right)$ is the
electrostatic Green's function
\begin{eqnarray}
&G\left( r_j, r_k\right)& = \sum_{n= -\infty}^{\infty}\left[
\frac{1}{\sqrt{\left( 2nL+x_k-x_j\right)^2 + \left(
y_k-y_j\right)^2}} \right.\nonumber \\
&& \left. \,\,\,\,\,\,\,\,\,\, \,\,\,\,\,\,\,\,\,\,
\,\,\,\,\,\,\,\,\,\, \,\,\,\,\,\,\,\,\,\,  -\frac{1}{\sqrt{\left(
2nL+x_k+x_j\right)^2 + \left( y_k-y_j\right)^2}} \right],
\label{eq:greenfunction}
\end{eqnarray}
which includes the effect of image charges representing the
screening effect of external electrodes modeled as ideally
conducting semi-spaces.

For practical calculations, we do not need to evaluate $U$ from
Eq.~(\ref{eq:totalsystemenergy}), because this equation may be used
to rewrite Eq.~(\ref{eq:systemenergydifference}) in the explicit
form
\begin{eqnarray}
\Delta U_{jk} &=& \varepsilon_j^{\left( 0 \right)} -
\varepsilon_k^{\left( 0 \right)} +e\mathbf{E}\mathbf{r}_{jk}
+ \frac{e^2}{\kappa} \sum_{l\neq j} \left(n_l-\frac{1}{2}\right) G\left( r_j, r_l\right) \nonumber \\
&& \,\,\,\,\,\,\,\,\,\,  -\frac{e^2}{\kappa} \sum_{l\neq k}
\left(n_l-\frac{1}{2}\right) G\left( r_k, r_l\right)+
\frac{e^2}{\kappa} G\left( r_j, r_k\right).
\label{eq:sitetositesystemenergydifference}
\end{eqnarray}

The numerical study has been carried out by using the classical
Monte Carlo technique based on the algorithm suggested by Bakhvalov
$\it {et~al.}$ \cite{Bakhvalovetal1989}, which has become the de
facto standard for single-electron tunneling simulations
\cite{Wasshuber}. In most cases, the calculated variables are
averaged over several (many) conductors with independent random
distributions of localized sites in space and energy, but the same
macroscopic parameters. The spectral density of current fluctuations
is calculated using the advanced algorithm described in detail in
Ref.~\cite{Method-KinkhabwalaSverdlovKorotkov2005}.

\section{\label{sec:level1}Results}

In order to classify the physical regimes of hopping behavior, it
is useful to note that our model has four relevant energy scales:

(i) $1/\nu_0 a^{2}$ describes the energy spectrum discreteness,

(ii) $eEa$ is the scale of the electric field
strength,

(iii) $e^2/\kappa a = \chi/\nu_0 a^{2}$ characterizes the Coulomb
interaction strength and

(iv) $k_{B}T$ is the scale of thermal fluctuations.

Our primary interest is in transport,
especially its dependence on the applied electric field $E$, so
that instead of comparing $eEa$ with the other three energy
scales, we prefer to speak about three characteristic values of
electric field,
which should be compared with the actual $E$:
\begin{equation}
eaE_T \equiv k_{B}T = \frac{T}{T_0} \times \frac{1}{\nu_0 a^2}, \\\ eaE_0\equiv \frac{1}{\nu_0 a^2}, \\\
eaE_{c}\equiv \frac{e^2}{\kappa a} = \chi \times \frac{1}{\nu_0
a^2}. \label{eq:characteristicfields}
\end{equation}
We are not interested in the case of extremely high temperatures, so
that we will always assume that $T \ll T_0$, i.e$.$ $E_T \ll E_0$.
On the other hand, the relative position of points $E_c$ and $E_0$
on the field axis is determined by the normalized parameter of the
Coulomb interaction strength:
\begin{equation}
E_c/E_0 = \chi \equiv e^2 \nu_0 a/ \kappa.
\label{eq:chi}
\end{equation}

\subsection{\label{sec:level1}Coulomb Gap}

In order to understand the peculiarities of Coulomb interaction
effects in our model, we started with a calculation of the
single-particle density of states for the case of $T = 0$ and $E =
0$. Indeed, in all the Coulomb gap analyses we are aware of, the
electrostatic boundary effects have been ignored by assuming
\begin{equation}
G\left( r_j, r_k\right) = \frac{1}{r_{jk}}.
\label{eq:bulkgreenfunction}
\end{equation}
On the contrary, in our Green's function (\ref{eq:greenfunction})
the image charge contribution may be substantial, so it has been
essential to understand how this contribution affects the Coulomb
gap formation.

Following the Coulomb gap literature
\cite{ShklovskiiBook,EfrosPollackCollection}, we define the
effective single-particle energy of an electron on site $j$ as
\begin{equation}
\varepsilon_j \equiv \varepsilon^{(0)}_j + \frac{e^2}{\kappa}
\sum_{l\neq j} (n_l-\frac{1}{2}) G\left( r_j, r_l\right).
\label{eq:singleparticleenergy}
\end{equation}
Note that our basic
Eq.~(\ref{eq:sitetositesystemenergydifference}) may be
conveniently rewritten in terms of $\varepsilon_j$:
\begin{equation}
\Delta U_{jk} = \varepsilon_j - \varepsilon_k
+e\mathbf{E}\mathbf{r}_{jk} + \frac{e^2}{\kappa} G\left( r_j,
r_k\right).
\label{eq:SingPartEnergysitetositesystemenergydifference}
\end{equation}

The calculations of the single-particle density of states in the
ground state of a system (in which all $\Delta U_{jk}$ are negative)
require its ``annealing". In our case the annealing is facilitated
by the fact that our model allows hopping between any pair of sites.
This is why the natural relaxation of the conductor at $T=0$ and
$E=0$ gave the results undistinguishable from those obtained after
an explicit annealing procedure (see, e.g.,
Ref.~\cite{KaplanSverdlovLikharev2003}). Since the used Monte Carlo
algorithm is not slowed down when all the transition rates are very
low, the relaxation could be simulated very quickly.

\begin{figure}
\begin{center}
\includegraphics[height=5.4in]{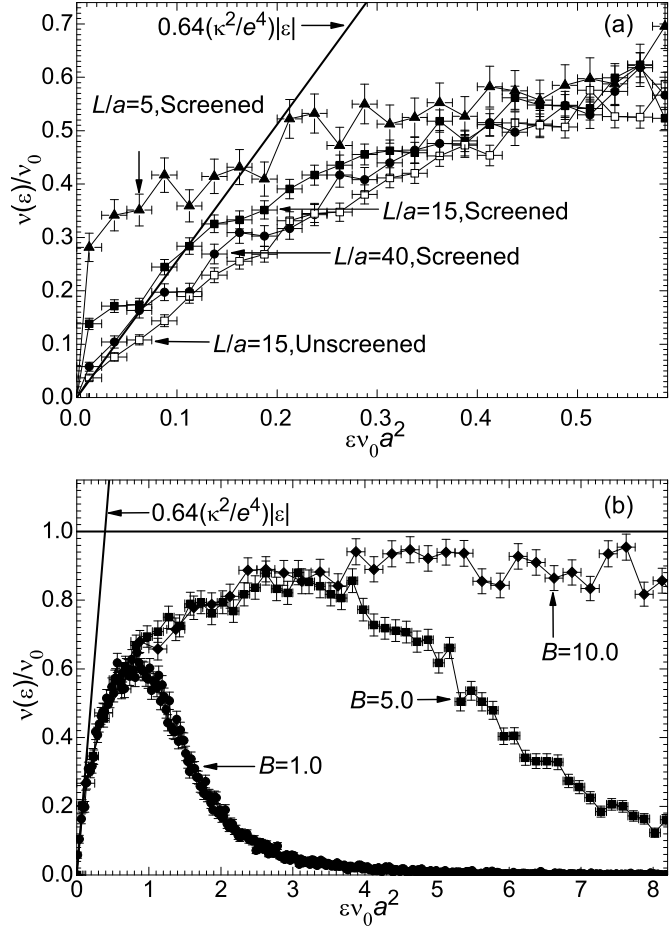}
\end{center}
\caption{Single-particle density of states $\nu\left( \varepsilon
\right)/\nu_0$ averaged over a large number of conductors at
$\chi=0.5$, $T=0$ and $E=0$ for (a) several conductor lengths $L$
for large width $W/a=40$ at fixed half-bandwidth of the seed energy
band ($B=1$) with and without the screening due to electrostatic
boundary effects, and for (b) several values of the half-bandwidth
$B$ for sufficiently large conductors with screening. The straight
lines correspond to Eq.~(\ref{eq:singleparticleDoS}) with
$\alpha=2/\pi \approx 0.64$. Curves are only guides for the eye.}
\label{fig:densityofstates}
\end{figure}

Figure~\ref{fig:densityofstates} shows our typical results for the
single-particle density of states. The soft Coulomb gap at
sufficiently low energies is clearly visible. The effects of
screening by the external electrodes are shown in
Fig.~\ref{fig:densityofstates}(a). The data labeled ``Screened"
correspond to the full Green's function (\ref{eq:greenfunction}),
which includes the electrostatic screening effects of the external
electrodes, while the results for the simple approximation
(\ref{eq:bulkgreenfunction}) are marked ``Unscreened". The results
show that for conductors of sufficiently large size, screening has
virtually no effect on the Coulomb gap formation. In this limit, the
linear part of the $\nu \left( \varepsilon \right)$ dependence
corresponds to Eq.~(\ref{eq:singleparticleDoS}) with the
self-consistent equation result $c = 2/\pi \approx 0.64$ cited
above.

Figure~\ref{fig:densityofstates}(b) shows the single-particle
density of states for three different values of an important
technical parameter, the half-bandwidth $B$ of the seed energy band.
One can see that the value of $B$ does not affect the
single-particle density of states well inside the Coulomb gap, but
may influence the results at larger energies, so that $B$ should
always be chosen properly in each particular case.

All the results presented below have been obtained for conductor
size $L \times W$ and energy bandwidth $2B$ so large that the
effects of screening and finite number of states are negligible.

\subsection{\label{sec:level1} DC Transport Characteristics}

\begin{figure}
\begin{center}
\includegraphics[height=3.4in]{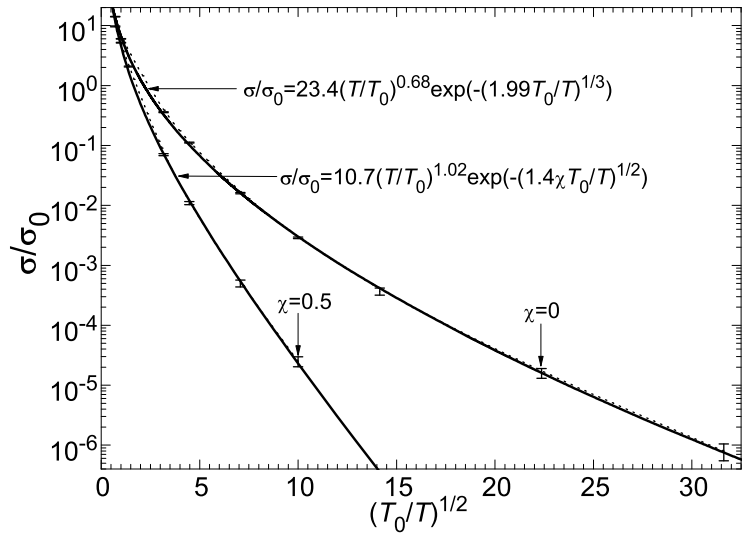}
\end{center}
\caption{Linear dc conductivity $\sigma$ for negligible Coulomb
interaction and finite Coulomb interaction as a function of
temperature. Points show the Monte Carlo results which were obtained
by the direct averaging of current calculated for a large number
(from $20$ to $96$) of conductors with a random distribution of
localized states, but the same macroscopic parameters. The sample
size ranged from $20 a \times 14 a$ to $80 a \times 50 a$, depending
on $\chi$ and $T$. Thin dashed lines are only guides for the eye,
while the thick solid lines correspond to the best fits of the data
by Eqs.~(\ref{eq:zerocoulombhightemperatureconductivity}) and
(\ref{eq:nonzerocoulombhightemperatureconductivity}).}
\label{fig:AllCoulombHighTemperatureConductivity}
\end{figure}

Figure~\ref{fig:AllCoulombHighTemperatureConductivity} shows our
Monte Carlo results for the dc conductivity $\sigma$ as a function
of temperature $T$ for two values of the Coulomb interaction
strength parameter $\chi$. The results for $\chi=0$ coincide with
those discussed in our previous work
\cite{2DKinkhabwalaSverdlovKorotkovLikharev2004}. In particular, for
sufficiently low temperatures ($E_T \ll E_0$), the Monte Carlo data
may be fit by Eq.~(\ref{eq:zerocoulombhightemperatureconductivity})
using a simple power law in temperature $T$ to estimate the
pre-exponential (model-dependent) function $A\left(T, 0, 0 \right)=
\left(23.4 \pm 1.3 \right)\left(T/T_0\right)^{\left( 0.68 \pm
0.04\right)}$ and a constant for $B\left(T,0,0 \right)=2.0 \pm 0.2$.
(See Ref.~\cite{2DKinkhabwalaSverdlovKorotkovLikharev2004} for a
detailed discussion of this result.)

On the other hand, the results for $\chi=0.5$ show that in the case
of substantial Coulomb interaction, the temperature dependence of
conductance at low temperatures ($T/T_0 \ll \chi^3$) follows the
Efros-Shklovskii variable-range hopping result
(\ref{eq:nonzerocoulombhightemperatureconductivity}). The best
fitting gives $A\left(T, 0, 0.5 \right)=\left(10.7 \pm 1.3
\right)\left(T/T_0\right)^{\left( 1.02 \pm 0.12\right)}$ and
$B\left(T, 0, 0.5 \right)=1.4 \pm 0.3$. This is in a reasonable
agreement with the following values for $B$ (in our units): $3.1$
following from an approximate analysis based on percolation theory
\cite{Nguen1984}, $4.8$ found by evaluating an approximate integral
over critical hops using a lattice model
\cite{LevinNguenShklovskiiEfros1987} and $2.9$ obtained for a
narrower range of temperatures using numerical (Monte Carlo)
simulations on a uniform periodic lattice with randomly distributed
energies \cite{TsigankovEfros2002}. (Unfortunately, the above values
had no uncertainty reported.) The difference between our result and
the reported values is probably due to the differences between
details of the used models - see Sec. 2 above.

For very high temperatures ($E_T \gtrsim E_0$), the exponential
temperature dependence of variable-range hopping theory cannot give
a good description of the results, because in this case transport is
dominated by very short hops with lengths of the order of the
localization radius. However, even at these temperatures, Coulomb
interaction effects lead to a drop in dc conductivity.

\begin{figure}
\begin{center}
\includegraphics[height=3.4in]{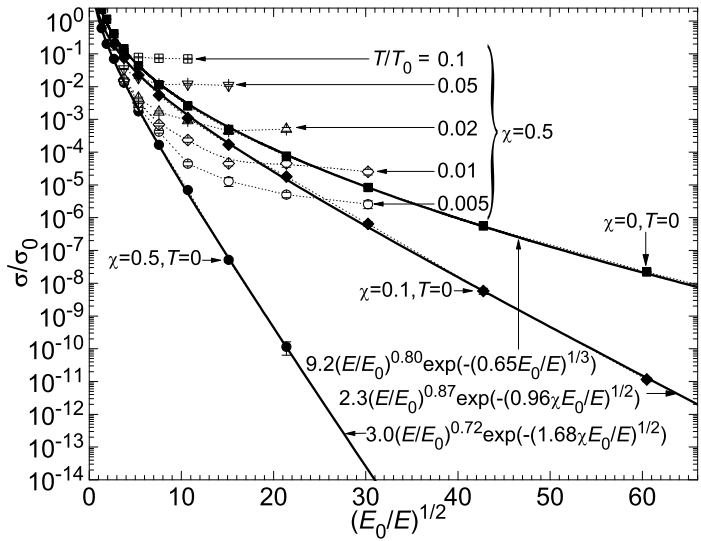}
\end{center}
\caption{Nonlinear dc conductivity $\sigma$ as a function of
electric field $E$ for several values of temperature $T$ and Coulomb
interaction strength $\chi$. Points are Monte Carlo results averaged
over a large number (from $20$ to $96$) of conductors of the same
size (ranging from $20 \times 14 a^2$ to $800 \times 500 a^2$,
depending on $\chi$, $T$ and $E$). Solid symbols show results for
$T=0$, while open symbols correspond to $T \neq 0$. Thin dashed
lines are only guides for the eye. Thick solid lines are the fits to
the $T=0$ results using
Eqs.~(\ref{eq:zerocoulombhighelectricfieldconductivity}) and
(\ref{eq:nonzerocoulombhighelectricfieldconductivity}).}
\label{fig:AllCandTandEconductivity}
\end{figure}

For higher electric fields ($E \gtrsim E_T$), the dc current
$\left\langle I \right\rangle$ increases faster than $E$, therefore
dc conductivity, defined as $\sigma \equiv \left\langle
I\right\rangle/E W$, becomes nonlinear and begins to increase with
$E$ - see Fig.~\ref{fig:AllCandTandEconductivity}. (More extensive
data for the case of negligible Coulomb interaction, $\chi=0$, are
shown in Fig.~2 of
Ref.~\cite{2DKinkhabwalaSverdlovKorotkovLikharev2004}.)

For $T\rightarrow 0$, our results
\cite{2DKinkhabwalaSverdlovKorotkovLikharev2004} for the case of
negligible Coulomb interaction ($\chi=0$) exhibit an exponential
field dependence for not very high fields $\left( E_T \ll E \ll E_0
\right)$ and may be fit for the entire range of fields considered by
Eq.~(\ref{eq:zerocoulombhighelectricfieldconductivity}) using
(analogous to the low field case) a simple power law in field $E$ to
estimate the pre-exponential (model-dependent) function
$A\left(0,E,0\right) = \left(9.2 \pm 0.6
\right)\left(E/E_0\right)^{\left( 0.80 \pm 0.02\right)}$ along with
constant $B\left(0,E,0 \right) = 0.65\pm 0.02$. (See
Ref.~\cite{2DKinkhabwalaSverdlovKorotkovLikharev2004} for a detailed
discussion of this result.) The corresponding results for $\chi =
0.1$ and $\chi = 0.5$ show that increasing Coulomb interaction
strength suppresses the nonlinear dc conductivity, just as in the
low-field case. These results may be well fit by
Eq.~(\ref{eq:nonzerocoulombhighelectricfieldconductivity}) with $A
\left( 0 , E , 0.1 \right) = \left(2.3 \pm 0.6
\right)\left(E/E_0\right)^{\left( 0.87 \pm 0.07\right)}$ and
$B\left(0,E,0.1 \right) = 0.96\pm 0.05$ for $\chi = 0.1$ and $A
\left( 0, E , 0.5 \right) = \left(3.0 \pm 0.4
\right)\left(E/E_0\right)^{\left( 0.72 \pm 0.07\right)}$ and
$B\left(0,E,0.5 \right) = 1.68\pm 0.07$ for $\chi=0.5$. It is
possible that any differences due to fitting reflect a (weak)
systematic dependence on $\chi$.

The results for very high electric fields, $E \gtrsim E_0$ (near the
localization limit), do  not obey the variable-range hopping theory,
due to very short hops of the order of the localization radius which
dominate the transport in this case. Note, however, that even within
this range the dc conductivity decreases with increasing Coulomb
interaction strength.

To summarize our dc transport results, we see a very reasonable
agreement with variable-range hopping theory within appropriate
parameter ranges. Moreover, we believe that our
supercomputer-enabled numerical modeling has given accurate
parameters for the coefficients of these theories for our particular
model.

\subsection{\label{sec:level1}Current Fluctuations}

\subsubsection{\label{sec:level1}$1/f$ Noise}

\begin{figure}
\begin{center}
\includegraphics[height=3.6in]{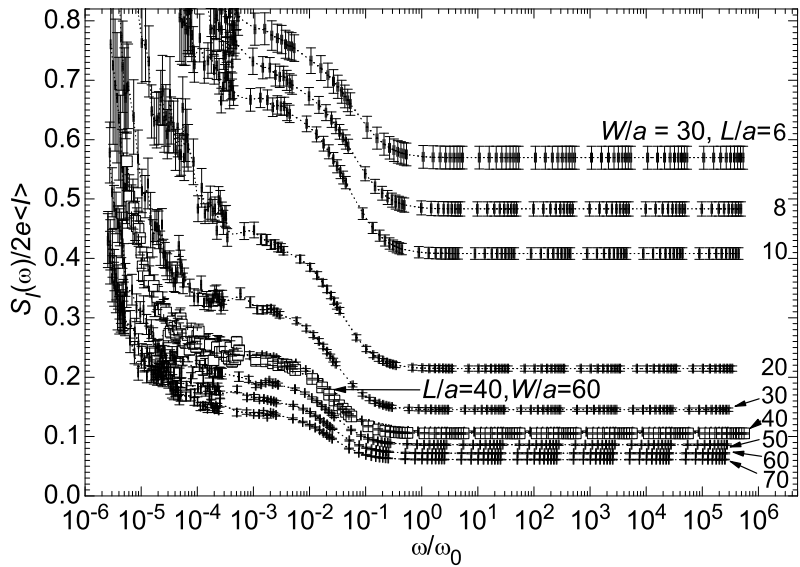}
\end{center}
\caption{Spectral density $S_I \left( \omega \right)$ of current
fluctuations at fixed Coulomb interaction strength $\chi=0.5$, as a
function of observation frequency $\omega$ measured in units of
$\omega_0 \equiv g/\hbar \nu_0 a^2$, for several values of conductor
length $L$. Each point represents data averaged over 48 conductor
samples at fixed parameters ($\chi=0.5$, $T=0$ and $E/E_0=0.07$).
Small points show results for $W/a=30$, while open squares are for
$W/a=60$ (at $L/a=40$). Thin dashed lines are only guides for the
eye.} \label{fig:C0.5specdensfrequencySchottky}
\end{figure}

Figure~\ref{fig:C0.5specdensfrequencySchottky} shows typical results
of our calculations of current noise at zero temperature, finite
Coulomb interaction strength and fixed electric field, for several
values of conductor length. Of particular note is that in sharp
contrast with the negligible Coulomb interaction case
\cite{2DKinkhabwalaSverdlovKorotkovLikharev2004}, we do observe a
$1/f$-type noise at $f \rightarrow 0$. The frequency $f_{k}$ of the
$1/f$ noise ``knee" (the crossover from this noise to a quasi-flat
spectral density) is relatively constant (or at most grows slowly
with decreasing conductor length). This is what could be expected
from the comparison of Eqs.~(\ref{eq:hoogenoisescaling}) and
(\ref{eq:fanolength}) if $\alpha \sim 1$. (For sufficiently large
conductor width $W$, $\left\langle I \right\rangle$ is proportional
to $W$, so that $f_k$ should also be independent of $W$ as well,
which is consistent with our results for different width in
Fig.~\ref{fig:C0.5specdensfrequencySchottky}.)

\begin{figure}
\begin{center}
\includegraphics[height=3.6in]{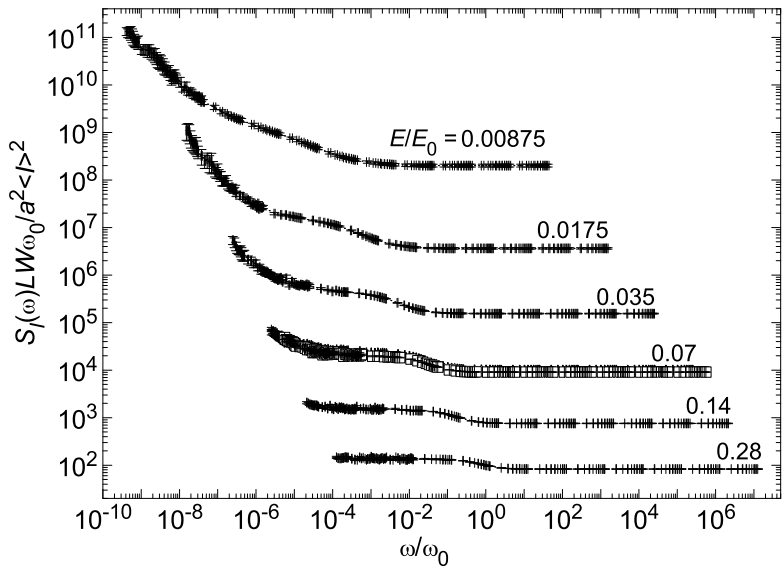}
\end{center}
\caption{Spectral density $S_I \left(\omega \right)$ of current
fluctuations at $T=0$ and $\chi=0.5$, normalized to the Hooge
scaling factor $a^2 \left< I\right>^2 /LW \omega_0$, as a function
of observation frequency $\omega$ (measured in units of $\omega_0$)
for several values of electric field. Each point represents data
averaged over 48 conductor samples of the same size (ranging from
$20 a\times 14 a$ to $120 a\times 60 a$, depending on $E$). Lines
are only guides for the eye. For $E/E_0=0.07$, the results are
plotted for a few conductor sizes, $50 a \times 30 a, 60 a \times 30
a \, \textrm{and} \, 70 a\times 30 a$ (small points) and $40 a
\times 60 a$ (open squares). The results imply that the $1/f$-type
noise (in this normalization) is virtually size- and
field-independent.} \label{fig:C0.5specdensfrequencyHooge}
\end{figure}

In Fig.~\ref{fig:C0.5specdensfrequencyHooge}, the calculation
results are plotted in the form allowing their straightforward
comparison with the Hooge scaling \cite{KoganBook,Hooge1969}. Indeed
in these coordinates, Eq.~(\ref{eq:hoogenoisescaling}) with $C(f)/f
\propto 1/f^p$ would give a straight line dropping with slope $p$.
We see that our data for $f \rightarrow 0$ are compatible with this
formula, with $p \sim 1.3$. This value is close to the one
calculated in Ref.~\cite{Reichhardt04} using a rather different
model of hopping, apparently more adequate near the metal-insulator
transition. The result is also compatible with recent experiments
\cite{Jaroszynski02} which indicate an increase of $p$ from
approximately 1 on the metallic side of such transition to above one
on its dielectric side.

Unfortunately, more accurate determination of the noise spectral
density $S_I(f)$ for sufficiently small frequencies and/or finite
temperatures has been out of our reach, despite the use of advanced
averaging algorithms
\cite{2DKinkhabwalaSverdlovKorotkovLikharev2004} and unique
supercomputer resources. As a result, at this stage we cannot
compare our results with the existing theories of $1/f$ noise at
hopping
\cite{Shklovskii1980_1981,Shklovskii2003,ShtengelYu2003,Kozub1996,
KozubBaranovskiiShlimak1999,Kogan1998}.

\subsubsection{\label{sec:level1}Fano Factor and Cluster Length}

\begin{figure}
\begin{center}
\includegraphics[height=5.4in]{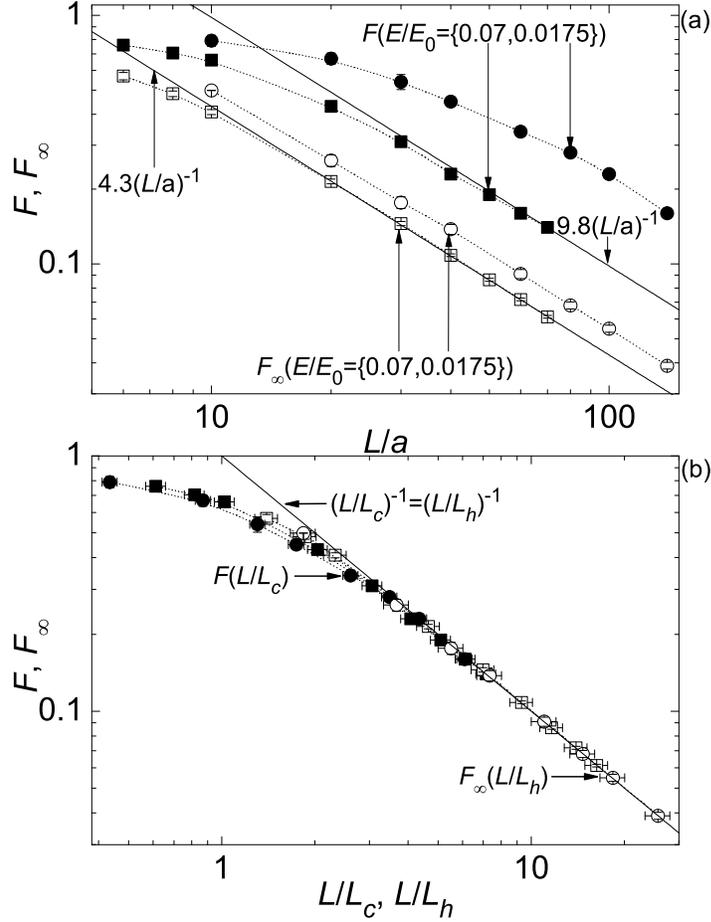}
\end{center}
\caption{Average Fano factor $F$ and its high-frequency counterpart
$F_{\infty}$ (Eqs.~(\ref{eq:fanofactoromegatoplateau}) and
(\ref{eq:inffreqfanofactor}), respectively) as functions of
conductor length $L$ normalized to: (a) the localization length $a$,
and (b) the scaling lengths $L_c$ (for $F$) and $L_h$ (for
$F_{\infty}$) (see Fig.~\ref{fig:clusterlengthelectricfield} below),
for two values of applied field at $\chi=0.5$, $T=0$ and $W \gg
L_c$. Straight lines are the best fits to the data (using
Eqs.~(\ref{eq:fanolength}) and
(\ref{eq:infinitefrequencyfanolength})), while dashed curves are
only guides for the eye.} \label{fig:fanoplateauvslength}
\end{figure}

If the low-frequency spectral density is constant (as it is for
hopping at negligible Coulomb interaction
\cite{1DKorotkovLikharev2000, 2DSverdlovKorotkovLikharev2001,
2DKinkhabwalaSverdlovKorotkovLikharev2004}), it is naturally
characterized by the Fano factor - see Eq.~(\ref{eq:fanofactor}). In
the presence of a $1/f$-type noise, the definition of the Fano
factor is less obvious. However,
Figs.~\ref{fig:C0.5specdensfrequencySchottky} and
\ref{fig:C0.5specdensfrequencyHooge} show that the fluctuation
spectrum has an exponentially broad plateau between the $1/f$ noise
knee and a crossover to another, high-frequency value. (The latter
crossover at higher frequencies exists even in the absence of the
Coulomb interaction - see
Ref.~\cite{2DKinkhabwalaSverdlovKorotkovLikharev2004} for a more
detailed discussion.) The large length of these plateaus gives a
motivation for the generalization of the Fano factor definition:
\begin{equation}
F \equiv \frac{S_I(f_{p})}{2e\left\langle I\right\rangle},
\label{eq:fanofactoromegatoplateau}
\end{equation}
where $f_{p}$ is any frequency between the $1/f$ noise knee and the
high frequency crossover. In addition, following
Ref.~\cite{2DKinkhabwalaSverdlovKorotkovLikharev2004}, we may define
a similar factor on the high frequency plateau:
\begin{equation}
F_{ \infty } \equiv \frac{S_I\left(f \rightarrow  \infty
\right)}{2e\left\langle I\right\rangle}.
\label{eq:inffreqfanofactor}
\end{equation}
Figure~\ref{fig:C0.5specdensfrequencySchottky} shows that neither of
these factors depend on the sample width, at least for reasonably
large $W$. Figure~\ref{fig:fanoplateauvslength} shows the dependence
of these factors on conductor length $L$. The results for $F$ in the
case of substantial Coulomb interactions agree well \cite{small_L}
with Eq.~(\ref{eq:fanolength}) with $\alpha \approx 1$, in contrast
with the result $\alpha \neq 1$ for negligible interaction
\cite{2DSverdlovKorotkovLikharev2001,
2DKinkhabwalaSverdlovKorotkovLikharev2004}. The results for the high
frequency counterpart $F_{ \infty }$ agree well with the similar
expression \cite{2DKinkhabwalaSverdlovKorotkovLikharev2004}
\begin{equation}
F_{\infty} = \left( \frac{L_h}{L}\right)^{\beta},\qquad L\gg L_h,
\label{eq:infinitefrequencyfanolength}
\end{equation}
where, within the accuracy of our calculations, $\beta = 1$. This
result is similar to that for negligible Coulomb interaction
\cite{2DKinkhabwalaSverdlovKorotkovLikharev2004}, and may be
interpreted as a result of ``capacitive division" of the discrete
increments of externally-measured charge jumps resulting from
single-electron hops through the system \cite{Av-Likh}. Figure
\ref{fig:fanoplateauvslength}(b) shows that both results can be
collapsed onto a universal scaling curve by the introduction of
certain length scales: $L_c$ for $F$ and $L_h$ for $F_{\infty}$.

\begin{figure}
\begin{center}
\includegraphics[height=3.6in]{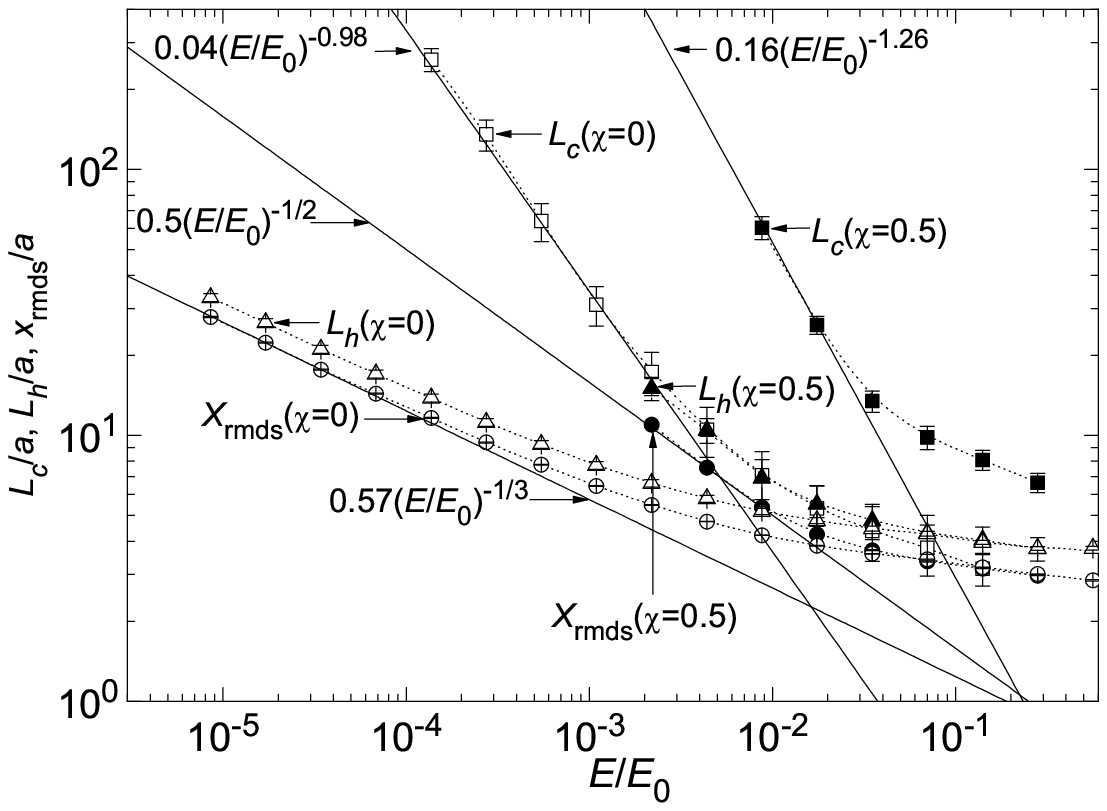}
\end{center}
\caption{The values of parameters $L_c$ and $L_h$ giving the best
fitting of shot noise results for Eqs.~(\ref{eq:fanolength}) and
(\ref{eq:infinitefrequencyfanolength}), respectively, for
sufficiently large conductors ($L,W \gg L_c$), as functions of
electric field at $T=0$ for the cases of negligible ($\chi=0$, open
squares and triangles) and substantial ($\chi=0.5$, solid squares
and triangles) Coulomb interaction. For comparison, circles show the
results for the simple direction-weighted average hop length along
the electric field direction (\ref{eq:xrmds}). Dashed curves are
only guides for the eye, while solid lines are the best fits using
the variable-range hopping and percolation theory predictions (see
the text).} \label{fig:clusterlengthelectricfield}
\end{figure}

In order to compare the corresponding length scales ($L_c$ and
$L_h$) with the proper measures of hop length, we have calculated
the direction-weighted average
\cite{2DKinkhabwalaSverdlovKorotkovLikharev2004} along the field
direction
\begin{equation}
x_{\mathrm{rmds}}^2 \equiv \frac{\sum_{j,k} x_{jk}^2\left\vert
H_{jk}-H_{kj}\right\vert}{\sum_{j,k} \left\vert
H_{jk}-H_{kj}\right\vert }, \label{eq:xrmds}
\end{equation}
where $x_{jk}\equiv x_k-x_j=-x_{kj}$ is the component of the $j
\rightarrow k$ hop length along the applied field direction, and
$H_{jk}$ is the number of electrons making this hop during a certain
time interval. For not too high fields ($E_T \ll E \ll E_0$), the
results in Fig.~\ref{fig:clusterlengthelectricfield} for negligible
Coulomb interaction are in good agreement with the variable-range
hopping scaling described by
Eq.~(\ref{eq:zerocoulombhighelectricfieldhoplength}), while for
substantial Coulomb interaction they follow scaling similar to
Eq.~(\ref{eq:nonzerocoulombhighelectricfieldhoplength}). In both
cases, $L_h$ and $x_{\mathrm{rmds}}$ have a similar behavior in the
entire range of studied fields. (See
Ref.~\cite{2DKinkhabwalaSverdlovKorotkovLikharev2004} for a more
detailed discussion on this result.) On the other hand, $L_c$, as
determined from Eq.~(\ref{eq:fanolength}), has a very different
scaling, especially for lower fields ($E \ll E_0$). Namely, at
negligible Coulomb interaction, the results for $L_c$ follow the law
(\ref{eq:clusterlengthstrongfield}) with $J=0.04\pm 0.01$ and
$\mu=0.98\pm 0.08$ \cite{2DKinkhabwalaSverdlovKorotkovLikharev2004}
(see section 1.3), while in the case of substantial Coulomb
interactions ($\chi = 0.5$), $L_c$ also obeys
Eq.~(\ref{eq:clusterlengthstrongfield}), but with $J=0.16\pm 0.02$
and $\mu=1.26\pm 0.15$.

Following the analysis of
Ref.~\cite{2DKinkhabwalaSverdlovKorotkovLikharev2004}, we may use
the theory of directed percolation
\cite{StaufferAharonyBook,Obukhov1980,Essametal1986} to predict the
following scaling:
\begin{equation}
L_c\propto \left\langle x\right\rangle \left( \frac{x_c}{\left\vert
\left\langle x\right\rangle-x_c\right\vert}
\right)^{\delta_{\parallel}},
\end{equation}
where $\left\langle x\right\rangle$ is approximately equal to
$x_{\mathrm{rmds}}$, $x_c$ is the critical hop length along the
field, and the critical index $\delta_{\parallel}$ should be close
\cite{Essametal1986} to $1.73$. Due to the exponential nature of the
variable-range hopping scaling, $\left\vert\left\langle
x\right\rangle - x_c\right\vert \sim a$, and depending on the regime
of hopping, we may use the corresponding field scaling of
Eqs.~(\ref{eq:zerocoulombhighelectricfieldhoplength}) or
(\ref{eq:nonzerocoulombhighelectricfieldhoplength}) (see
Fig.~\ref{fig:clusterlengthelectricfield}) to arrive at
Eq.~(\ref{eq:clusterlengthstrongfield}) with either
$\mu=\frac{1}{3}\left( 1+\delta_{\parallel}\right)\approx 0.91$ or
$\mu=\frac{1}{2}\left( 1+\delta_{\parallel}\right)\approx 1.37$,
respectively. These values are in a very reasonable agreement with
our simulation results, thus confirming the interpretation of $L_c$
as the average length of the directed percolation cluster.

In extremely high fields ($E \gtrsim E_0$), the length scales $L_c$
and $x_{\mathrm{rmds}}$ become comparable to one another and both
approach the localization radius $a$.

\section{\label{sec:level1}Discussion}

To summarize, we have carried out numerical simulations of 2D
hopping within a broad range of temperature, electric field and
Coulomb interaction strength. For average (dc) transport
characteristics, our results are in general agreement with the
variable-range hopping theories, except for the (model-dependent)
cases of ``ultra-high" electric field and/or temperature, where the
hopping length becomes of the order of the localization radius.

For the spectrum of current fluctuations, our results are more
significant. First, for the case of significant Coulomb interaction
we have obtained a reliable evidence of $1/f$-like fluctuations,
approximately obeying the Hooge scaling
(\ref{eq:hoogenoisescaling}), even at $T \rightarrow 0$. In
hindsight, this result does not seem too surprising. Due to the
presence of Coulomb interaction, random motion of the electrons
during hopping transport generates a time- and space-varying Coulomb
field, with a quasi-white spectrum, even at $T = 0$. The effect of
such a randomly changing field on localized electrons aside from
from the hopping clusters should be qualitatively similar to that of
thermal fluctuations that may lead to the $1/f$ noise, for example
following one of the scenarios described in
Refs.~\cite{Shklovskii1980_1981, Shklovskii2003, ShtengelYu2003,
Kozub1996, KozubBaranovskiiShlimak1999}. Recent experiments for
hopping in quasi-3D samples \cite{Shlimak1995,Raquet1999}, showing a
very slow change of $1/f$ noise intensity at $T \rightarrow 0$, seem
qualitatively compatible with this interpretation.

Our second important result is that in the presence of significant
Coulomb interaction, the quasi-white noise above the $1/f$ noise
knee is suppressed according to the scaling law
(\ref{eq:fanolength}) with $\alpha = 1$ (within the accuracy of our
numerical experiment). This result is consistent with the simple
addition of mutually-independent noise voltages generated by
(conductor) sample sections connected in series, and hence with the
existence of a finite correlation length. On the other hand, the
results \cite{2DKinkhabwalaSverdlovKorotkovLikharev2004} for
negligible Coulomb interactions give $\alpha=0.76\pm 0.08 < 1$, and
are inconsistent with the existence of such length, at least on the
scale of $L_c$. However, in both cases the constant $L_c$,
participating in the scaling law~(\ref{eq:fanolength}), may be
interpreted as the length between the critical hops, i.e$.$ the
directed percolation cluster length.

\ack{Fruitful discussions with A. Efros, S. Kogan, A. Korotkov, V.
Kozub, M. Pollak, C. Reichhart, B. Shklovskii and D. Tsigankov are
gratefully acknowledged. The work was supported in part by the
Engineering Physics Program of the Office of Basic Energy Sciences
at the U.S. Department of Energy, and by the Semiconductor Research
Corporation. We also acknowledge the use of the following
supercomputer resources: SBU's cluster $Njal$ (purchase and
installation funded by U.S. DoD's DURINT program), Oak Ridge
National Laboratory's IBM SP computer $Eagle $ (funded by the
Department of Energy's Office of Science and Energy Efficiency
program), and also IBM SP system $Tempest$ at Maui High Performance
Computing Center and IBM SP system $Habu$ at NAVO Shared Resource
Center (computer time granted by DOD's High Performance Computing
Modernization Program).}

\end{document}